\documentclass[10pt,twoside]{article}
\usepackage[applemac]{inputenc}
\usepackage{amssymb,amsmath}
\usepackage{graphicx}

\usepackage{pstricks}
\usepackage{pst-plot}
\usepackage{pst-math}
\def\volnum{second draft}
\usepackage{aflbcours2015}
\pdebut{1}

\title{Electric charge in hyperbolic motion: The early history and other geometrical aspects}
\titleshort{Electric charge in hyperbolic motion \dots}
\author{C\u alin Galeriu}
\authorshort{C. Galeriu}
\address{Becker College \\ 
61 Sever Street, Worcester, MA 01609, USA}

\begin{document}
\maketitle

\vskip 1cm
\begin{abstract}
R\'ESUM\'E. Nous revisitons les premiers travaux de Minkowski et Sommerfeld concernant le mouvement hyperbolique, et nous d\'ecrivons certains aspects g\'eom\'etriques de l'interaction \'electrodynamique. Nous discutons les avantages d'une formulation sym\'etrique en temps dans laquelle les points mat\'eriels sont remplac\'es par des \'el\'ements de longueur infinit\'esimale.

{\it
ABSTRACT. We revisit the early work of Minkowski and Sommerfeld concerning hyperbolic motion, and we describe some geometrical aspects of the electrodynamic interaction. We discuss the advantages of a time symmetric formulation in which the material points are replaced by infinitesimal length elements.}
\end{abstract}

\section{A Review of Minkowski's Work}

The description of an electric charge in hyperbolic motion is at the center of Minkowski's geometrical formulation of electrodynamics. In "Space and Time" \cite{minkowski}, his last contribution to special relativity before his premature death, Minkowski gives a brief geometrical recipe for calculating the four-force with which an electric charge acts upon another electric charge. We can get a better appreciation of Minkowski's revolutionary work by exploring his geometrical recipe, by assembling the derivation details that Minkowski did not have a chance to publish.

The first step in Minkowski's procedure is to approximate the motion of the source charge with hyperbolic motion. This is done with the help of three infinitesimally close points on the worldline of the source particle. This hyperbola of curvature, tangent to the worldline, is determined by the four-velocity and the four-acceleration of the particle at the given spacetime point $P$. The hyperbola lies in the plane defined by these two four-vectors. The four-acceleration, which is perpendicular to the four-velocity, lies on a line that goes through the center of the hyperbola $M$. Minkowski mentions that the length $\rho$ of the segment MP determines the magnitude of the four-acceleration, according to the formula $c^2 / \rho$.

\begin{figure}[h!]
\begin{center}
  \includegraphics[height=8cm]{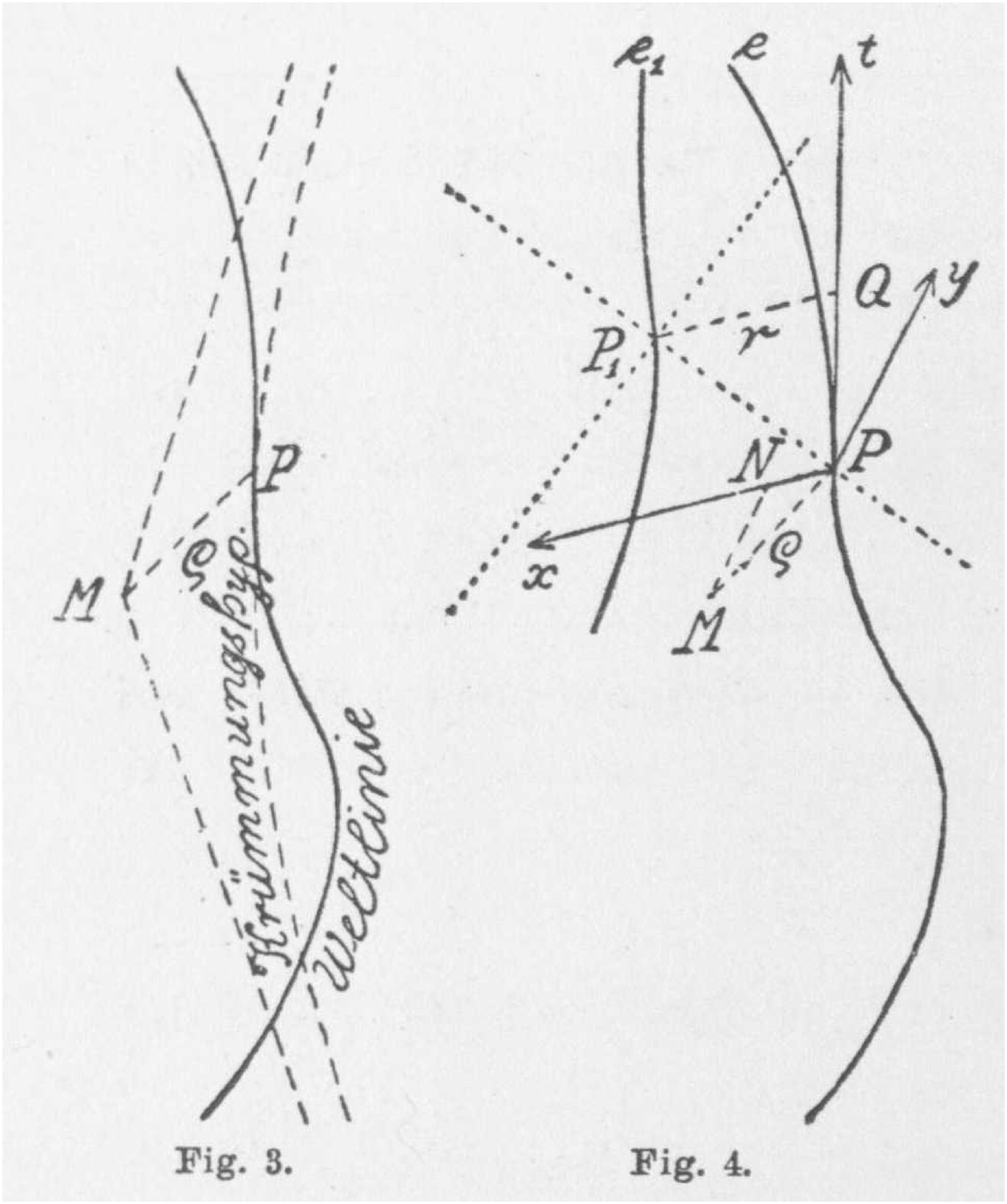}
  \caption{Minkowski's original diagrams from "Space and Time".}
  \label{fig:minkowski}
\end{center}
\end{figure}

The second step in Minkowski's procedure is to complete the spacetime diagram with the inclusion of the test charge. From the worldpoint $P$, where the source particle is, we draw into the future a lightcone that intersects the worldline of the test particle at point $P_1$. Minkowski uses a reference frame which is instantaneously at rest with the source particle at $P$. In this co-moving reference frame the time axis $Pt$ has the direction of the four-velocity, which is tangent to the worldline. The distance $r$ between the test charge and the source charge, at the retarded position, is the length of a segment $P_1Q$ perpendicular to the time axis $Pt$. In the co-moving reference frame the four-potential, being parallel to the four-velocity of the source particle, has an especially simple expression. Minkowski mentions that the four-potential at worldpoint $P_1$ has the direction of $PQ$ and the magnitude of $e / r$ (in Gaussian units), where $e$ is the electric charge of the source.

The third step in Minkowski's procedure is to establish the spatial axes of the co-moving reference frame. The origin of the reference system is at $P$, the retarded position of the source charge. The $x$ axis is parallel to the radial segment $QP_1$, previously determined. The $y$ axis is chosen in such a way to make sure that the hyperbola of curvature (the plane $(MPQ)$) and the field point $P_1$ are included in the three dimensional subspace spanned by the $Px$, $Py$, and $Pt$ axes. The $y$ axis is parallel to the segment $MN$, which is drawn from $M$, the center of the hyperbola, perpendicular to the $Px$ axis. It turns out that the segment $MN$ is not only perpendicular to the $Px$ axis, but also perpendicular to the $Pt$ axis, which makes it eligible to indicate the direction of the $y$ axis. (The proof that $MN \perp PQ$ goes like this: $PQ \perp MP$ because the four-velocity is perpendicular to the four-acceleration. $PQ \perp NP$ because $PQ \perp P_1Q$ and $NP \parallel P_1Q$. As a result $PQ \perp (MPN)$, which means that $PQ \perp MN$.) The $z$ axis is automatically determined from its orthogonality to the $x$, $y$, and $t$ axes. 

In the final step Minkowski writes down an expression for the four-force with which the particle at $P$ (with electric charge $e$), acts upon the particle at $P_1$ (with electric charge $e_1$). The electrodynamic four-force (in Gaussian units) is:  

\begin{equation}
\mathbf{F} = - e e_1 (\dot{t}_1 - \frac{\dot{x}_1}{c}) \mathbf{K},
\label{eq:1}
\end{equation}
where the dots represent derivation with respect to the proper time. The components of the four-vector $\mathbf{K}$ satisfy the relations:
\begin{equation}
c K_t - K_x = \frac{1}{r^2}, \ \ \ \ \ 
K_y = \frac{\ddot{y}}{c^2 r}, \ \ \ \ \ 
K_z = 0,
\label{eq:2}
\end{equation}
while, at the same time, the four-vector $\mathbf{K}$ is orthogonal to the four-velocity of the test particle at $P_1$. We also notice that in Minkowski space $u = i c t$ and $K_u = i c K_t$.

It turns out that, in order to fully understand the geometrical recipe of Minkowski, one first has to derive an expression of the electromagnetic field produced by the source charge. This is a step that Minkowski has probably intentionally left out. 

Born \cite{born} is the first one to publish a calculation of the electromagnetic field produced by an electric charge distribution in hyperbolic motion. He uses these results in his discussion of the stability of the rigid electron. Sommerfeld \cite{sommerfeld} soon realizes that, for all practical purposes, an electron acts just like a point particle, and gives a detailed derivation of the electromagnetic field produced by a point charge in hyperbolic motion. Laue \cite{laue} reproduces Sommerfeld's derivation in his book, giving even more mathematical details. Unfortunately, Laue fails to follow up with a similar calculation of the electrodynamic four-force. One decade later the derivation of the electromagnetic field of a point charge in hyperbolic motion briefly resurfaces in Pauli's \cite{pauli} book.  

With the exception of Sommerfeld \cite{sommerfeld}, who mentions that his own calculation of the electrodynamic four-force is in agreement with the geometric rule given by Minkowski, the early special relativity authors have completely ignored Minkowski's geometrical description of the interaction between two electrically charged particles. As an example, even Pauli \cite{pauli} fails to mention that his algebraic derivation of the electrodynamic four-force reproduces Minkowski's expression in all the details. This important validation of Minkowski's geometrical approach did not escape Walter \cite{walter}, who mentions it, still without proof, in footnote (111) of a recent essay. We want to bring to light all the derivation details behind the results of Minkowski, Sommerfeld, and Pauli, and to present the early history of hyperbolic motion in a comprehensive and fully accessible manner.

\section{A Review of Sommerfeld's Work}

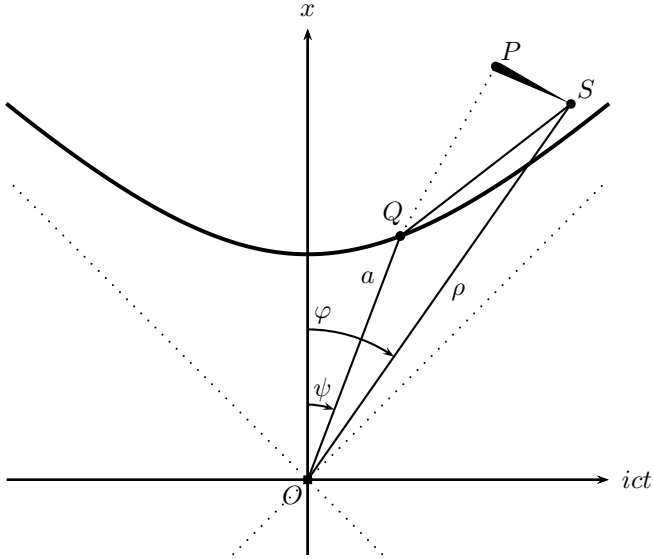
\begin{figure}[h!]
\begin{center}
\begin{pspicture}(-4,-1)(4,6.5)
\psaxes[ticks=none,labels=none]{->}(0,0)(-4,-1)(4,6)[$i c t$,0][$x$,90]
\rput(-0.2,-0.2){$O$}
\psline[linestyle=dotted](-1,-1)(4,4)
\psline[linestyle=dotted](1,-1)(-4,4)
\psparametricplot[plotstyle=curve,linewidth=1.5pt]{-1.1}{1.1}{t SINH 3 mul t COSH 3 mul} %a=3
\psline[showpoints=true](0,0)(1.232,3.243) %t=0.4
\rput(1.132,3.543){$Q$}
\psarc{<-}(0,0){1}{69.2}{90}
\rput(0.2,1.2){$\psi$}
\psline[showpoints=true](0,0)(3.5,5)
\rput(3.7,5.2){$S$} 
\psline(1.232,3.243)(3.5,5)
\psarc{<-}(0,0){2}{55.0}{90}
\rput(0.2,2.2){$\varphi$}
\psline[showpoints=true,linestyle=dotted](1.232,3.243)(2.5,5.5)
\rput(2.7,5.7){$P$}
\pspolygon*(3.5,5)(2.45,5.45)(2.55,5.55)
\rput(0.8,2.7){$a$}
\rput(2.0,2.5){$\rho$}
\end{pspicture}
\caption{The source particle at $Q$ exerts a force on the test particle at $P$. Point $S$ is the projection of point $P$ onto the plane of the hyperbola, and $PS \perp QS$. We notice that $PS^2 = y^2 + z^2$, $QS^2 = a^2 + \rho^2 - 2 a \rho \cos(\psi - \varphi)$, and $QP^2 = PS^2 + QS^2 = 0$.}
\label{fig:hyperbola}
\end{center}
\end{figure}

A point in Minkowski space is described by a position four-vector $\mathbf{X} = (x, y, z, i c t)$. Consider a particle in hyperbolic motion, moving along the $x$ axis, as shown in figure \ref{fig:hyperbola}. The center of the hyperbola $O$ is taken as the origin of the reference frame. If the length of the segment $OQ$ is $a$, and if the angle between this segment and the $Ox$ axis is $\psi$, then a point $Q$ on the worldline of the particle will have coordinates $\mathbf{X}_Q = (a \cos(\psi), 0, 0, a \sin(\psi))$. The arc length $s$ on the hyperbola is given by $s = a~\psi$. The infinitesimal arc length is related to the proper time of the particle by $ds = i~c~d\tau$. The particle at $Q$ will have a four-velocity $\mathbf{V}_Q = (-i c \sin(\psi), 0, 0, i c \cos(\psi))$ and a four-acceleration $\mathbf{A}_Q = (\frac{c^2}{a} \cos(\psi), 0, 0, \frac{c^2}{a} \sin(\psi))$. The magnitudes of these four-vectors are $|\mathbf{X}_Q| = a$, $|\mathbf{V}_Q| = i c$, $|\mathbf{A}_Q| = \frac{c^2}{a}$. The reference frame in which the particle at Q is instantaneously at rest (the four-velocity is parallel to the time axis) comes from a rotation of the original axes by an angle $\psi$: 
\begin{equation}
\begin{pmatrix}A_x' \\[0.3em] A_u'\end{pmatrix} = 
\begin{pmatrix}\cos(\psi) & \sin(\psi) \\[0.3em] 
-\sin(\psi) \text{  }& \cos(\psi)\end{pmatrix}
\begin{pmatrix}\frac{c^2}{a} \cos(\psi) \\[0.3em] \frac{c^2}{a} \sin(\psi)\end{pmatrix} = 
\begin{pmatrix}\frac{c^2}{a} \\[0.3em] 0\end{pmatrix}.
\label{eq:rotation}
\end{equation}

In the co-moving reference frame the acceleration experienced by the particle is always equal to $\frac{c^2}{a}$. This is the acceleration presented by Minkowski in the first step of his procedure given in "Space and Time". This acceleration, constant in the proper reference frame, provides an alternative and equivalent way of defining hyperbolic motion. 

We want to determine the electrodynamic four-force due to an electric charge $e$ in hyperbolic motion, acting on an electric charge $e_1$. The source particle is at point $Q$, while the test particle is at point $P$ with coordinates $\mathbf{X}_P = (\rho \cos(\varphi), y, z, \rho \sin(\varphi))$. Because worldpoints $P$ and $Q$ are connected by a light signal, the four-vector $\mathbf{X} = \mathbf{X}_P - \mathbf{X}_Q$, from the source charge to the test charge, must have a magnitude of zero. 
(While Pauli works with a four-vector $\mathbf{X}$ as defined here, Sommerfeld works with a four-vector $\mathbf{R} = \mathbf{X}_Q - \mathbf{X}_P$. To compare formulas one has to replace $\mathbf{R}$ with $-\mathbf{X}$.) Since $\mathbf{X} = (\rho \cos(\varphi) - a \cos(\psi), y, z, \rho \sin(\varphi) - a \sin(\psi))$, the condition $\mathbf{X} \cdot \mathbf{X} = 0$ reduces to the equation:
\begin{equation}
\cos(\psi - \varphi) = \frac{\rho^2 + y^2 + z^2 + a^2}{2 a \rho}.
\label{eq:3}
\end{equation}

The independent variables are $\rho, y, z, \varphi$ (the coordinates of the test particle), and equation (\ref{eq:3}) expresses the implicit dependence of the position of the retarded source particle on the position of the test particle, $\psi = \psi(\rho, y, z, \varphi)$. By partial differentiation of equation (\ref{eq:3}) we obtain some relations needed later for the calculation of the electric field:
\begin{equation}
\frac{\partial \psi}{\partial \rho} = \frac{a \cos(\psi - \varphi) - \rho}{a \rho \sin(\psi - \varphi)},
\label{eq:4}
\end{equation}
\begin{equation}
\frac{\partial \psi}{\partial y} = \frac{- y}{a \rho \sin(\psi - \varphi)},
\label{eq:5}
\end{equation}
\begin{equation}
\frac{\partial \psi}{\partial z} = \frac{- z}{a \rho \sin(\psi - \varphi)}.
\label{eq:6}
\end{equation}

Laue \cite{laue} also mentions that the angle $\psi - \varphi$ is imaginary and negative. This is a good time to calculate some scalar products:
\begin{equation}
\mathbf{V}_Q \cdot \mathbf{X} = - i c \rho \sin(\psi - \varphi),
\label{eq:VX}
\end{equation}
\begin{equation}
\mathbf{A}_Q \cdot \mathbf{X} = \frac{c^2 \rho}{a} \cos(\psi - \varphi) - c^2.
\label{eq:AX}
\end{equation}

The retarded four-potential at point $P$ (in Gaussian units) is:
\begin{equation}
\Phi = - \frac{e \mathbf{V}_Q}{\mathbf{V}_Q \cdot \mathbf{X}}.
\label{eq:7}
\end{equation}

Sommerfeld and Pauli, using rational electrostatic units, have a extra $4 \pi$ in the denominator. Using (\ref{eq:VX}), the four-potential becomes:
\begin{equation}
\Phi = \left( \frac{- e}{\rho}\frac{\sin(\psi)}{\sin(\psi - \varphi)}, 0, 0, \frac{e}{\rho}\frac{\cos(\psi)}{\sin(\psi - \varphi)} \right).
\label{eq:8}
\end{equation}

A rotation of the $x$ and $t$ axes by an angle $\psi$ brings us to a reference frame in which the source charge at $Q$ is instantaneously at rest. The components of the four-potential become:
\begin{equation}
\Phi ' = \left( 0, 0, 0, \frac{e}{\rho}\frac{1}{\sin(\psi - \varphi)} \right).
\label{eq:9}
\end{equation}

This is the four-potential presented by Minkowski in the second step of his procedure given in "Space and Time". We will prove that:
\begin{equation}
\rho \sin(\psi - \varphi) = - i r.
\label{eq:radius}
\end{equation}

\begin{figure}[h!]
\begin{center}
\begin{pspicture}(-4,-1)(4,6.5)
\psaxes[ticks=none,labels=none]{->}(0,0)(-4,-1)(4,6)[$i c t$,0][$x$,90]
\rput(-0.2,-0.2){$O$}
\psline[linestyle=dotted](-1,-1)(4,4)
\psline[linestyle=dotted](1,-1)(-4,4)
\psparametricplot[plotstyle=curve,linewidth=1.5pt]{-1.1}{1.1}{t SINH 3 mul t COSH 3 mul}
\psline[showpoints=true](0,0)(-1.232,3.243)
\rput(-1.432,3.043){$Q$}
\psarc(0,0){0.5}{90}{110.8}
\rput(-0.12,0.7){$\theta$}
\psline(-4,4.294)(4,1.255)
\psline[showpoints=true](0,2.775)(0,5)
\rput(-0.2,2.575){$U$}
\rput(0.2,5.2){$S$}
\pspolygon*(0,5)(-1.05,5.45)(-0.95,5.55)
\rput(-1.2,5.7){$P$}
\psline[showpoints=true](-1,5.5)(0.9877,2.3998)
\rput(1.1877,2.5998){$T$}
\psline(0,5)(0.9877,2.3998)
\psarc(0,5){0.5}{-90}{-69.2}
\rput(0.12,4.3){$\theta$}
\psline[linestyle=dotted](-1.232,3.243)(-1,5.5)
% line QT has equation y = -0.37986 x + 2.775
% line ST has equation y = -2.63252 x + 5
% angle theta is 20.8 degrees
\end{pspicture}
\caption{In the co-moving reference frame of the source particle at $Q$ the test particle at $P$ is described by a radial position vector $TP$.}
\label{fig:radius}
\end{center}
\end{figure}

We work in a reference frame in which the field point and the center of the hyperbola are simultaneous, as shown in figure \ref{fig:radius}. The source particle at $Q$ has a four-velocity tangent to the hyperbola. We draw this tangent line. The projection of point $S$ on this tangent line is point $T$. Because $QT \perp ST$ and $PS \perp QT$ we conclude that $QT \perp (PST)$, and as a result $PT \perp QT$. We identify $PT$ as the segment $P_1Q$ of length $r$ from Minkowski's "Space and Time". The intersection of line $QT$ with line $OS$ is point $U$. The segment $OQ$ has length $a$, the segment $OS$ has length $\rho$, and the segment $PS$ has length $\sqrt{y^2 + z^2}$. By definition $\theta = \varphi - \psi$, this is an imaginary and positive angle. Care should be taken when comparing our formulas with Sommerfeld's, because he uses an imaginary and negative angle instead.

We notice that we have two similar right triangles, $\Delta OQU \sim \Delta STU$. Because $QO \perp QU$ we have $OU = \frac{OQ}{\cos(\theta)} = \frac{a}{\cos(\theta)}$. By subtraction $US = OS - OU = \rho - \frac{a}{\cos(\theta)}$. Because $TU \perp TS$ we have $TS = US \cos(\theta) = \rho \cos(\theta) - a$. Using equation (\ref{eq:3}) we are able to write:
\begin{multline}
r^2 = PT^2 = TS^2 + PS^2 = [\rho \cos(\theta) - a]^2 + y^2 + z^2 \\
= \rho^2 \cos^2(\theta) - 2 \rho a \cos(\theta) + a^2 + y^2 + z^2 \\
= \rho^2 \cos^2(\theta) - (\rho^2 + y^2 + z^2 + a^2) + a^2 + y^2 + z^2 \\
= -\rho^2 [1 - \cos^2(\theta)] = -\rho^2 \sin^2(\theta).
\label{eq:radius2}
\end{multline}

Keeping the positive solution, $r = -i \rho \sin(\theta) = i \rho \sin(\psi - \varphi)$. q.e.d.

We now return to equation (\ref{eq:8}). A rotation of the $x$ and $t$ axes by an angle $\varphi$ brings us to a reference frame in which the field point $P$ and the center of the hyperbola $O$ are simultaneous. The components of the four-potential become:
\begin{equation}
\Phi '' = \left( \frac{- e}{\rho}, 0, 0, \frac{e}{\rho}\frac{\cos(\psi - \varphi)}{\sin(\psi - \varphi)} \right).
\label{eq:10}
\end{equation}

The electric and magnetic fields are identified as elements in the electromagnetic field tensor matrix. The trick is to calculate $\mathcal{F}_{ik} = \frac{\partial \Phi_k}{\partial x_i} - \frac{\partial \Phi_i}{\partial x_k}$ as a rotor (curl) in cylindrical coordinates \cite{laue}. The calculation is done in the reference frame in which the field point and the center of the hyperbola are simultaneous. In this reference system the $\rho$ axis is parallel to the $x$ axis, the $\varphi$ axis is parallel to the $u = i c t$ axis, and 
\begin{equation}
(\Phi_\rho, \Phi_y, \Phi_z, \Phi_\varphi) = \left( \frac{- e}{\rho}, 0, 0, \frac{e}{\rho}\frac{\cos(\psi - \varphi)}{\sin(\psi - \varphi)} \right).
\label{eq:11}
\end{equation}

Keep in mind that $\Phi_y = \Phi_z = 0$ and that $\Phi_\rho$ depends only on $\rho$. The magnetic and electric fields are given by the equations:
\begin{equation}
H_x = \mathcal{F}_{yz} = \frac{\partial \Phi_z}{\partial y} - \frac{\partial \Phi_y}{\partial z} = 0,
\label{eq:12}
\end{equation}
\begin{equation}
H_y = \mathcal{F}_{zx} = \mathcal{F}_{z\rho} = \frac{\partial \Phi_\rho}{\partial z} - \frac{\partial \Phi_z}{\partial \rho} = 0,
\label{eq:13}
\end{equation}
\begin{equation}
H_z = \mathcal{F}_{xy} = \mathcal{F}_{\rho y} = \frac{\partial \Phi_y}{\partial \rho} - \frac{\partial \Phi_\rho}{\partial y} = 0,
\label{eq:14}
\end{equation}
\begin{equation}
- i E_x = \mathcal{F}_{xu} = \mathcal{F}_{\rho \varphi} = \frac{1}{\rho} \frac{\partial (\rho \Phi_\varphi)}{\partial \rho} - \frac{1}{\rho} \frac{\partial \Phi_\rho}{\partial \varphi} = \frac{1}{\rho} \frac{\partial (\rho \Phi_\varphi)}{\partial \rho},
\label{eq:15}
\end{equation}
\begin{equation}
- i E_y = \mathcal{F}_{yu} = \mathcal{F}_{y \varphi} = \frac{\partial \Phi_\varphi}{\partial y} - \frac{1}{\rho} \frac{\partial \Phi_y}{\partial \varphi} = \frac{\partial \Phi_\varphi}{\partial y}, 
\label{eq:16}
\end{equation}
\begin{equation}
- i E_z = \mathcal{F}_{zu} = \mathcal{F}_{z \varphi} = \frac{\partial \Phi_\varphi}{\partial z} - \frac{1}{\rho} \frac{\partial \Phi_z}{\partial \varphi} = \frac{\partial \Phi_\varphi}{\partial z}. 
\label{eq:17}
\end{equation}

With the help of (\ref{eq:4})-(\ref{eq:6}), equations (\ref{eq:15})-(\ref{eq:17}) become \cite{pauli}:
\begin{equation}
E_x = \frac{- i e}{\rho \sin^2(\psi - \varphi)} \frac{\partial \psi}{\partial \rho} = \frac{- i e [ a \cos(\psi - \varphi) - \rho ]}{a \rho^2 \sin^3(\psi - \varphi)},
\label{eq:18}
\end{equation}
\begin{equation}
E_y = \frac{- i e}{\rho \sin^2(\psi - \varphi)} \frac{\partial \psi}{\partial y} = \frac{i e y}{a \rho^2 \sin^3(\psi - \varphi)},
\label{eq:19}
\end{equation}
\begin{equation}
E_z = \frac{- i e}{4 \pi \rho \sin^2(\psi - \varphi)} \frac{\partial \psi}{\partial z} = \frac{i e z}{4 \pi a \rho^2 \sin^3(\psi - \varphi)}.
\label{eq:20}
\end{equation}

Having derived these results Sommerfeld \cite{sommerfeld} states that "These formulas are the most simple expressions of the field produced by any moving point charge." However some simplifications can still be made by substituting $\rho \sin(\psi - \varphi)$ with $- i r$. In this way we get:
\begin{equation}
E_x = \frac{e \rho [\rho - a \cos(\psi - \varphi)]}{a r^3},
\label{eq:21}
\end{equation}
\begin{equation}
E_y = \frac{e \rho y}{a r^3},
\label{eq:22}
\end{equation}
\begin{equation}
E_z = \frac{e \rho z}{a r^3}.
\label{eq:23}
\end{equation}

Using equations (\ref{eq:3}) and (\ref{eq:radius}) we are able to write:
\begin{multline}
[\rho - a \cos(\psi - \varphi)]^2 + y^2 + z^2 \\
= \rho^2 - 2 \rho a \cos(\psi - \varphi) + a^2 \cos^2(\psi - \varphi) + y^2 + z^2 \\
= \rho^2 - (\rho^2 + y^2 + z^2 + a^2) + a^2 \cos^2(\psi - \varphi) + y^2 + z^2 \\
= -a^2 [1 - \cos^2(\psi - \varphi)] = - a^2 \sin^2(\psi - \varphi) = \frac{r^2 a^2}{\rho^2}.
\label{eq:24}
\end{multline}

As a result of (\ref{eq:24}) the magnitude of the electric field is:
\begin{equation}
|\vec{E}| = \sqrt{E_x^2 + E_y^2 + E_z^2} = \frac{e}{r^2}.
\label{eq:25}
\end{equation}

This extremely simple result should not come as a surprise, since $\vec{H}^2 - \vec{E}^2$ is a fundamental invariant of the electromagnetic field tensor. Because $\vec{H} = 0$, $\vec{E}^2$ has to be a Lorentz invariant, and it is, given the fact that $\mathbf{V}_Q \cdot \mathbf{X} = - c r$. Sommerfeld does not mention it, but only because $\vec{H} = 0$ we can say that $\mathbf{E} = (E_x, E_y, E_z, 0)$ is a four-vector.

The direction of the electric field is given by the vector $(\rho - a \cos(\psi - \varphi), y, z)$. We will prove that this is the vector $MP$ in figure \ref{fig:electric}.

\begin{figure}[h!]
\begin{center}
\begin{pspicture}(-4,-1)(4,6.5)
\psaxes[ticks=none,labels=none]{->}(0,0)(-4,-1)(4,6)[$i c t$,0][$x$,90]
\rput(-0.2,-0.2){$O$}
\psline[linestyle=dotted](-1,-1)(4,4)
\psline[linestyle=dotted](1,-1)(-4,4)
\psparametricplot[plotstyle=curve,linewidth=1.5pt]{-1.1}{1.1}{t SINH 3 mul t COSH 3 mul}
\psline[showpoints=true](0,0)(-1.232,3.243)
\rput(-1.432,3.043){$Q$}
\psarc(0,0){0.5}{90}{110.8}
\rput(-0.12,0.7){$\theta$}
\psline(-4,4.294)(4,1.255)
\psline[showpoints=true](0,2.775)(0,5)
\rput(-0.2,2.575){$U$}
\rput(0.2,5.2){$S$}
\pspolygon*(0,5)(-1.05,5.45)(-0.95,5.55)
\rput(-1.2,5.7){$P$}
\psline(-1.232,3.243)(0,3.243)
\psline[showpoints=true](0,3.243)(1.232,3.243)
\rput(0.22,3.443){$M$}
\rput(1.432,3.043){$N$}
\psline[linestyle=dotted](-1.232,3.243)(-1,5.5)
\psline[linestyle=dotted](1.232,3.243)(-1,5.5)
\psline[showpoints=true](0,3.243)(-1,5.5)
\end{pspicture}
\caption{In the reference frame in which the test particle at $P$ and the center of the hyperbola $O$ are simultaneous the electric field has the direction of vector $MP$.}
\label{fig:electric}
\end{center}
\end{figure}

The projection of point $Q$ onto the $x$ axis is point $M$. The segment $OQ$ has length $a$, and the segment $OS$ has length $\rho$. Because $QM \perp OM$ we have $OM = OQ \cos(\theta) = a \cos(\theta)$ and $QM = OQ \sin(\theta) = a \sin(\theta)$. By subtraction $MS = OS - OM = \rho - a \cos(\theta)$. q.e.d.

For our point particle with electric charge $e_1$ and four-velocity $\mathbf{U} = (U_x, U_y, U_z, U_u)$ the electrodynamic four-force $\mathbf{F}$ (in Gaussian units) is:
\begin{equation}
\mathbf{F} = \frac{e_1}{c} 
\begin{pmatrix} 0 & 0 & 0 & -i E_x \\
0 & 0 & 0 & -i E_y \\
0 & 0 & 0 & -i E_z \\
i E_x & i E_y & i E_z & 0 \end{pmatrix}
\begin{pmatrix} U_x \\ U_y \\ U_z \\ U_u \end{pmatrix} 
= \frac{e_1}{c} 
\begin{pmatrix} - i E_x U_u \\ - i E_y U_u \\ - i E_z U_u \\
i (E_x U_x + E_y U_y + E_z U_z) \end{pmatrix}.
\label{eq:26}
\end{equation}

Sommerfeld introduces a unit four-vector in the temporal direction, $\mathbf{I} = (0, 0, 0, i)$. With the help of this four-vector the expression (\ref{eq:26}) becomes:
\begin{equation}
\mathbf{F} = - \frac{e_1}{c} ( \mathbf{U} \cdot \mathbf{I} ) \mathbf{E} + \frac{e_1}{c} ( \mathbf{U} \cdot \mathbf{E} ) \mathbf{I}.
\label{eq:27}
\end{equation}

Sommerfeld decomposes the direction of the four-vector $\mathbf{E}$ as:
\begin{equation}
\overrightarrow{MP} = \overrightarrow{QP} - \overrightarrow{QM} = \mathbf{X} - \overrightarrow{QM}. 
\label{eq:sommerfeld}
\end{equation}

The four-vector of the electric field $\mathbf{E} = (E_x, E_y, E_z, 0)$ is:
\begin{equation}
\mathbf{E} = \frac{e}{r^2} \frac{\overrightarrow{MP}}{MP} = \frac{e}{r^2} \frac{\mathbf{X} - \overrightarrow{QM}}{MP} = \frac{e \rho}{a r^3} (\mathbf{X} - \overrightarrow{QM}).
\label{eq:28}
\end{equation}

In this way the expression (\ref{eq:27}) of the four-force becomes:
\begin{equation}
\mathbf{F} = - \frac{e e_1 \rho}{c a r^3} ( \mathbf{U} \cdot \mathbf{I} ) (\mathbf{X} - \overrightarrow{QM})
 + \frac{e e_1 \rho}{c a r^3} ( \mathbf{U} \cdot (\mathbf{X} - \overrightarrow{QM}) ) \mathbf{I}.
\label{eq:29}
\end{equation}

Because $\overrightarrow{QM} = \frac{a \sin(\theta)}{i} \mathbf{I}$, the terms with $\overrightarrow{QM}$ in (\ref{eq:29}) cancel, and we are left with:
\begin{equation}
\mathbf{F} = - \frac{e e_1 \rho}{c a r^3} ( \mathbf{U} \cdot \mathbf{I} ) \mathbf{X} + \frac{e e_1 \rho}{c a r^3} ( \mathbf{U} \cdot \mathbf{X}) \mathbf{I}.
\label{eq:30}
\end{equation}

The final step is to write (\ref{eq:30}) in an explicitly invariant form, using only four-vectors and their scalar products. 

In the reference frame in which the test particle and the center of the hyperbola are simultaneous the source charge at $Q$ has a four-velocity:
\begin{equation}
\mathbf{V}_Q'' = i c (- \sin(\psi - \varphi), 0, 0, \cos(\psi - \varphi)) = i c ( \sin(\theta), 0, 0, \cos(\theta)),
\label{eq:31}
\end{equation}
and a four-acceleration:
\begin{equation}
\mathbf{A}_Q'' = \frac{c^2}{a} (\cos(\psi - \varphi), 0, 0, \sin(\psi - \varphi)) = \frac{c^2}{a} (\cos(\theta), 0, 0, - \sin(\theta)).
\label{eq:32}
\end{equation}

We can write the four-vector $\mathbf{I}$ as a linear combination of the orthogonal four-vectors $\mathbf{V}_Q''$ and $\mathbf{A}_Q''$. To save space we drop the subscript $Q$. The equation will be good in any reference frame, and there is no need to keep the $''$ on the four-vectors.
\begin{equation}
\mathbf{I} = \frac{(\mathbf{I} \cdot \mathbf{V})}{V^2} \mathbf{V} + \frac{(\mathbf{I} \cdot \mathbf{A})}{A^2} \mathbf{A} = \frac{\cos(\theta)}{c} \mathbf{V} + \frac{-i a \sin(\theta)}{c^2} \mathbf{A}
\label{eq:33}
\end{equation}

From equations (\ref{eq:VX}) and (\ref{eq:AX}) we have:
\begin{equation}
\sin(\theta) = \frac{\mathbf{V} \cdot \mathbf{X}}{i c \rho},
\label{eq:34}
\end{equation}
\begin{equation}
\cos(\theta) = \frac{a}{\rho} \frac{c^2 + \mathbf{A} \cdot \mathbf{X}}{c^2}.
\label{eq:35}
\end{equation}

Substituting (\ref{eq:34}) and (\ref{eq:35}) into (\ref{eq:33}) we get:
\begin{equation}
\mathbf{I} = \frac{a}{\rho} \frac{c^2 + (\mathbf{A} \cdot \mathbf{X})}{c^3} \mathbf{V} - \frac{a}{\rho} \frac{(\mathbf{V} \cdot \mathbf{X})}{c^3} \mathbf{A}.
\label{eq:36}
\end{equation}

Substituting (\ref{eq:36}) into (\ref{eq:30}), and using the fact that $\mathbf{V} \cdot \mathbf{X} = - c r$, brings the electrodynamic four-force to the final formula: 
\begin{multline}
\mathbf{F} = \frac{e e_1}{c (\mathbf{V} \cdot \mathbf{X})^3}  [ (c^2 + (\mathbf{A} \cdot \mathbf{X})) ( \mathbf{U} \cdot \mathbf{V} ) - (\mathbf{V} \cdot \mathbf{X}) (\mathbf{U} \cdot \mathbf{A}) ] \mathbf{X} \\
- \frac{e e_1}{c (\mathbf{V} \cdot \mathbf{X})^3} (c^2 + (\mathbf{A} \cdot \mathbf{X})) ( \mathbf{U} \cdot \mathbf{X} ) \mathbf{V} \\
+ \frac{e e_1}{c (\mathbf{V} \cdot \mathbf{X})^2} (\mathbf{U} \cdot \mathbf{X}) \mathbf{A}.
\label{eq:37}
\end{multline}

\section{The Equivalence of Minkowski's and Sommerfeld's Formulations}

The first step when comparing Minkowski's and Sommerfeld's expressions of the electrodynamic four-force is to perform a rotation of Sommerfeld's reference frame, in the $yOz$ plane, such that the $z$ coordinate of the field point $P$ vanishes. The fact that $z = 0$ in Minkowski's presentation is the reason why $K_z = 0$. Indeed, from (\ref{eq:23}) we get $E_z = 0$.

The fact that the four-force (\ref{eq:37}) is orthogonal to the four velocity $\mathbf{U}$ can easily verified by computing the scalar product $\mathbf{F} \cdot \mathbf{U}$. This orthogonality is of course a consequence of using an antisymmetric electromagnetic field tensor in the calculation of the four-force (\ref{eq:26}).

Looking at figure \ref{fig:minkowski}, and using Minkowski's reference frame and notation, we identify the four-vectors of interest as:
\begin{equation}
\mathbf{X} = (r, 0, 0, i r),
\label{eq:38}
\end{equation}
\begin{equation}
\mathbf{V} = (0, 0, 0, i c),
\label{eq:39}
\end{equation}
\begin{equation}
\mathbf{A} = (\ddot{x}, \ddot{y}, 0, 0),
\label{eq:40}
\end{equation}
\begin{equation}
\mathbf{U} = (\dot{x}_1, \dot{y}_1, \dot{z}_1, i c \dot{t}_1),
\label{eq:41}
\end{equation}
where the dots represent differentiation with respect to the proper time. With these four-vectors we calculate the scalar products of interest as:
\begin{equation}
\mathbf{X} \cdot \mathbf{V} = - r c,
\label{eq:42}
\end{equation}
\begin{equation}
\mathbf{X} \cdot \mathbf{A} = r \ddot{x},
\label{eq:43}
\end{equation}
\begin{equation}
\mathbf{X} \cdot \mathbf{U} = r \dot{x}_1 - r c \dot{t}_1,
\label{eq:44}
\end{equation}
\begin{equation}
\mathbf{U} \cdot \mathbf{V} = - c^2 \dot{t}_1,
\label{eq:45}
\end{equation}
\begin{equation}
\mathbf{U} \cdot \mathbf{A} = \ddot{x} \dot{x}_1 + \ddot{y} \dot{y}_1.
\label{eq:46}
\end{equation}

Substitution of the four-vectors (\ref{eq:38})-(\ref{eq:40}) and of the scalar products (\ref{eq:42})-(\ref{eq:46}) into (\ref{eq:37}) gives:
\begin{multline}
\mathbf{F} = \frac{e e_1}{c (- r c)^3}  [ (c^2 + r \ddot{x}) (- c^2 \dot{t}_1) + r c (\ddot{x} \dot{x}_1 + \ddot{y} \dot{y}_1) ] (r, 0, 0, i r) \\
- \frac{e e_1}{c (- r c)^3} (c^2 + r \ddot{x}) (r \dot{x}_1 - r c \dot{t}_1) (0, 0, 0, i c) \\
+ \frac{e e_1}{c (- r c)^2} (r \dot{x}_1 - r c \dot{t}_1) (\ddot{x}, \ddot{y}, 0, 0).
\label{eq:47}
\end{multline}

In Minkowski's reference frame $\mathbf{X}$ and $\mathbf{V}$ have a zero $y$ component. The third row of (\ref{eq:47}) gives the $y$ component of the four-force:
\begin{equation}
F_y = \frac{e e_1}{r c^3} (\dot{x}_1 - c \dot{t}_1) \ddot{y},
\label{eq:48}
\end{equation}
which matches Minkowski's expression from (\ref{eq:1}) and (\ref{eq:2}) perfectly. 

Due to the fact that in Minkowski's reference frame the $x$ and $u$ components of $\mathbf{X}$ are related by a factor of $i$, when calculating $\frac{F_u}{i} - F_x$ the contributions from the first row of (\ref{eq:47}) cancel out. What is left is:
\begin{equation}
\frac{F_u}{i} - F_x = \frac{e e_1}{r^2 c^3} (c^2 + r \ddot{x}) (\dot{x}_1 - c \dot{t}_1) - \frac{e e_1}{r c^3} (\dot{x}_1 - c \dot{t}_1) \ddot{x}, 
\label{eq:49}
\end{equation}
which, after some simplifications, again matches Minkowski's expression from (\ref{eq:1}) and (\ref{eq:2}) perfectly. We notice that Minkowski's claim of "full simplicity" for his description is a little bit exaggerated. Written in full, the $F_x$ and $F_u$ components are not as simple as Minkowski wants us to believe.

\section{A Review of Pauli's Work}

Pauli \cite{pauli} gives an algebraic derivation of the electrodynamic four-force. Let $X_{Qi}(\tau)$ describe the worldline of the source charge $Q$ as a function of its proper time $\tau$. The condition for a light signal:
\begin{equation}
X_i X_i = (X_{Pi} - X_{Qi})(X_{Pi} - X_{Qi}) = 0,
\label{eq:51}
\end{equation}
determines $\tau$ as a unique function of the four $X_{Pi}$ coordinates. By partial derivation of (\ref{eq:51}) with respect to $X_{Pk}$ we get:
\begin{equation}
X_i \left( \frac{\partial X_Pi}{\partial X_{Pk}} - \frac{\partial X_{Qi}}{\partial \tau} \frac{\partial \tau}{\partial X_{Pk}} \right) = 0.
\label{eq:52}
\end{equation}

Since $\frac{\partial X_Pi}{\partial X_{Pk}} = \delta_{ik}$ and $\frac{\partial X_{Qi}}{\partial \tau} = V_i$, equation (\ref{eq:52}) gives:
\begin{equation}
\frac{\partial \tau}{\partial X_{Pk}} = \frac{X_k}{(X_r V_r)},
\label{eq:53}
\end{equation}
and, as a consequence:
\begin{equation}
\frac{\partial X_i}{\partial X_{Pk}} = \frac{\partial X_Pi}{\partial X_{Pk}} - \frac{\partial X_{Qi}}{\partial \tau} \frac{\partial \tau}{\partial X_{Pk}} = \delta_{ik} - V_i \frac{X_k}{(X_r V_r)},
\label{eq:54}
\end{equation}
\begin{equation}
\frac{\partial V_i}{\partial X_{Pk}} = \frac{\partial V_i}{\partial \tau} \frac{\partial \tau}{\partial X_{Pk}} = 
A_i \frac{X_k}{(X_r V_r)}.
\label{eq:55}
\end{equation}

The retarded four-potential (in Gaussian units) is:
\begin{equation}
\Phi_i = - \frac{e V_i}{(V_r X_r)}.
\label{eq:50}
\end{equation}

The electromagnetic field  tensor at point $P$ is calculated as:
\begin{equation}
\mathcal{F}_{ik} = \frac{\partial \Phi_k}{\partial X_{Pi}} - \frac{\partial \Phi_i}{\partial X_{Pk}}.
\label{eq:56}
\end{equation}

Straightforward application of the rules of calculus, with the help of equations (\ref{eq:54})-(\ref{eq:55}) and $V_r V_r = -c^2$, gives \cite{pauli}:
\begin{multline}
\mathcal{F}_{ik} = - \frac{e}{(X_r V_r)^3} (c^2 + (X_r A_r)) (V_i X_k - V_k X_i) \\
+ \frac{e}{(X_r V_r)^2} (A_i X_k - A_k X_i).
\label{eq:57}
\end{multline}

The electrodynamic four-force is calculated as \cite{pauli}:
\begin{multline}
F_i = \frac{e_1}{c} \mathcal{F}_{ik} U_k \\
= - \frac{e e_1}{c (X_r V_r)^3} (c^2 + (X_r A_r)) (V_i (X_k U_k) - (V_k U_k) X_i) \\
+ \frac{e e_1}{c (X_r V_r)^2} (A_i (X_k U_k) - (A_k U_k) X_i),
\label{eq:58}
\end{multline}
an expression that is identical to the four-force (\ref{eq:37}).

It is no wonder that the physicists of last century have largely ignored the geometrical recipe of Minkowski and the geometrical derivation of Sommerfeld. The algebraic derivation of Pauli is much shorter and a lot easier to understand. It seems likely that Sommerfeld himself has some trouble with the use of Minkowski diagrams, drawing a two dimensional diagram instead of a three dimensional one. This is like making the implicit assumption that the test charge is in the same plane as the hyperbola. Sommerfeld also works with $(i E_x, i E_y, i E_z, 0)$ and $(0, 0, 0, 1)$ instead of $(E_x, E_y, E_z, 0)$ and $(0, 0, 0, i)$, not realizing that by definition a four-vector has three real and one imaginary components. And at one point he writes that the angle between the orthogonal four-velocity and four-acceleration vectors is $\frac{\pi}{2}$. We hope that our detailed geometrical derivations here and in \cite{galeriu3D} will make the use of 3D Minkowski diagrams less confusing, and that the original groundbreaking work of Minkowski and Sommerfeld will not be forgotten.

\section{The Geometrical Argument in Favor of Time Symmetric Electrodynamics}

We are now ready for a moment of reflection. How is it possible that the only effect that the acceleration of the source particle has on the electric field produced is a change in direction, while the magnitude stays the same? What is the process by which the electric field ends up with the direction of segment $MP$? Point $P$ is of course selected as the point where the electric field acts, but what is the mechanism by which point $M$ is selected? 

A careful look at figure \ref{fig:electric} shows that point $M$ is the midpoint of segment $QN$, where $Q$ is the position of the retarded source charge and $N$ is the position of the advanced source charge. We argue that point $M$ defines the direction of the electric field because the electrodynamic interaction is time symmetric. Instead of decomposing the segment $MP$ as in (\ref{eq:sommerfeld}) we can write:
\begin{equation}
\overrightarrow{MP} = \frac{1}{2} (\overrightarrow{QP} + \overrightarrow{NP}).
\label{eq:galeriu}
\end{equation}

This geometrical description of the electric field is so amazingly simple and beautiful, it brings to life Minkowski's vision: "[...] physical laws might find their most perfect expression as reciprocal relations between these worldlines." \cite{minkowski} 

The same time symmetric geometrical description applies to the case of a source charge in uniform motion \cite{galeriuAAD}. When the source charge moves with constant velocity, or with constant acceleration, our geometrical description is identical to the generally accepted causal theory. Since most of classical electrodynamics is concerned with such motion of the particles, one could hardly find any experimental evidence against (\ref{eq:galeriu}). Even if the electric charges had a variable acceleration, that would not be reflected in the electromagnetic field produced. The variable acceleration shows up only in the expression of the radiative damping force, and that is still a controversial topic \cite{galeriuRR}. In classical electrodynamics the instantaneous radiation reaction force is outside of direct experimental investigation.

Other authors have embraced a time symmetric formulation of electrodynamics. The absorber theory of radiation of Wheeler and Feynman \cite{wheelerfeynman1, wheelerfeynman2}, consistent with variational principles and conservation laws, has eliminated the divergence brought by self-interacting electrons, and has provided a derivation of the radiative damping force. The advanced converging waves predicted are practically impossible to be detected in a macroscopic experiment \cite{costa2}. Time symmetric interactions have been invoked by Costa de Beauregard \cite{costa3, costa} to explain retrocausality (the EPR paradox) in quantum mechanics, and by Woodward \cite{woodward} to explain the gravitational origin (Mach principle) of inertial reaction forces. Synge \cite{synge} has shown that if the retarded and advanced potentials have equal weights, and if the magnitudes of the acceleration of the source charge at the retarded and advanced positions are the same, then the electric charge does not radiate energy. While Synge discusses the case of a particle in uniform circular motion, the same conclusion applies to a particle in hyperbolic motion. This is a very significant result, because whether an electric charge in hyperbolic motion does or does not radiate energy is the subject of a long scientific controversy \cite{fabri}. Pauli \cite{pauli} predicts no radiation, based on the nul magnetic field (Poynting vector). Feynman \cite{feynman} also predicts no radiation, based on the principle of equivalence of general relativity. 

\section{The Geometrical Argument Against the Material Point Particle Model}

The time symmetric formulation of electrodynamics has an additional benefit: it can provide a geometrical explanation of a very puzzling fact. Suppose that we have a test charge $q$ in an electric field $\vec{E}$, with no magnetic field present. The force acting on the test charge is $\vec{F} = q \vec{E}$. If the velocity $\vec{v}$ of the particle is zero, then its four-velocity is $(\vec{0}, i c)$ and the four-force is $(\vec{F}, 0)$. But if the velocity $\vec{v}$ is not zero, then its four-velocity is $(\gamma(v) \vec{v}, i \gamma(v) c)$ and the four-force is $(\gamma(v) \vec{F}, i \frac{\gamma(v)}{c} \vec{F} \cdot \vec{v} )$. Why is the four-force dependent on the velocity of the test charge? If a particle is just a fixed point in Minkowski space, then it has no velocity! 

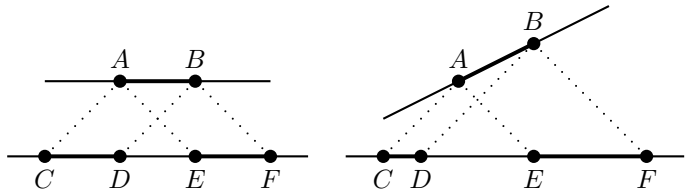
\begin{figure}[h!]
\begin{center}
\begin{pspicture}(-4.5,-0.5)(4.5,2)
\psline(-4.5,0)(-0.5,0)
\psline[showpoints=true,linewidth=1.5pt](-4,0)(-3,0)
\rput(-4,-0.3){$C$}
\rput(-3,-0.3){$D$}
\psline[showpoints=true,linewidth=1.5pt](-2,0)(-1,0)
\rput(-2,-0.3){$E$}
\rput(-1,-0.3){$F$}
\psline[linestyle=dotted](-4,0)(-3,1)
\psline[linestyle=dotted](-3,0)(-2,1)
\psline[linestyle=dotted](-2,0)(-3,1)
\psline[linestyle=dotted](-1,0)(-2,1)
\psline(-4,1)(-1,1)
\psline[showpoints=true,linewidth=1.5pt](-3,1)(-2,1)
\rput(-3,1.3){$A$}
\rput(-2,1.3){$B$}
\psline(0,0)(4.5,0)
\psline[showpoints=true,linewidth=1.5pt](0.5,0)(1,0)
\rput(0.5,-0.3){$C$}
\rput(1,-0.3){$D$}
\psline[showpoints=true,linewidth=1.5pt](2.5,0)(4,0)
\rput(2.5,-0.3){$E$}
\rput(4,-0.3){$F$}
\psline[linestyle=dotted](0.5,0)(1.5,1)
\psline[linestyle=dotted](1,0)(2.5,1.5)
\psline[linestyle=dotted](2.5,0)(1.5,1)
\psline[linestyle=dotted](4,0)(2.5,1.5)
\psline(0.5,0.5)(3.5,2)
\psline[showpoints=true,linewidth=1.5pt](1.5,1)(2.5,1.5)
\rput(1.5,1.3){$A$}
\rput(2.5,1.8){$B$}
\end{pspicture}
\caption{A simple case where the source particle is at rest, at the origin, and the test particle has only a radial velocity, if any. The test particle, segment $AB$, interacts with a retarded segment $CD$ and with an advanced segment $EF$.}
\label{fig:TSAAD}
\end{center}
\end{figure}

To solve this puzzle we have assumed that particles are not points in Minkowski space, but infinitesimal length elements along their worldlines \cite{galeriuAAD}. The electrodynamic interaction takes place between small worldline segments that have their start and end points connected by light signals, as shown in figure \ref{fig:TSAAD}. LaMont \cite{lamont} has experienced a similar insight, for he talks about the "infinitesimal thickness" of the lightcone of the source charge and "how much proper time the [test] charge spends in this region". 

Consider two points, $\mathbf{X}_1$ (on the source particle worldline) and $\mathbf{X}_2$ (on the test particle worldline) connected by a light signal. Consider two other points infinitely close to the first two points, $\mathbf{X}_1 + \mathbf{\delta X}_1$ (on the source particle worldline) and $\mathbf{X}_2 + \mathbf{\delta X}_2$ (on the test particle worldline), also connected by a light signal. Since $(\mathbf{X}_1 - \mathbf{X}_2) \cdot (\mathbf{X}_1 - \mathbf{X}_2) = 0$, $(\mathbf{X}_1 + \delta \mathbf{X}_1 - \mathbf{X}_2 - \delta \mathbf{X}_2) \cdot (\mathbf{X}_1 + \delta \mathbf{X}_1 - \mathbf{X}_2 - \delta \mathbf{X}_2) = 0$, $\delta \mathbf{X}_1 = \mathbf{V}_1 \delta \tau_1$, $\delta \mathbf{X}_2 = \mathbf{V}_2 \delta \tau_2$, to first order in the infinitesimals the ratio of the two infinitesimal segments connected by light signals at both ends is \cite{lamont}:
\begin{equation}
\frac{i c \delta \tau_1}{i c \delta \tau_2} = \frac{(\mathbf{X}_1 - \mathbf{X}_2) \cdot \mathbf{V}_2}{(\mathbf{X}_1 - \mathbf{X}_2) \cdot \mathbf{V}_1}.
\label{eq:lamont}
\end{equation}

When the velocity $\vec{v}$ of the test particle is zero:
\begin{equation}
\frac{i c \delta \tau_1^{(0)}}{i c \delta \tau_2} = \frac{(\mathbf{X}_1 - \mathbf{X}_2) \cdot (\vec{0}, i c)}{(\mathbf{X}_1 - \mathbf{X}_2) \cdot \mathbf{V}_1},
\label{eq:59}
\end{equation}
otherwise:
\begin{equation}
\frac{i c \delta \tau_1}{i c \delta \tau_2} = \frac{(\mathbf{X}_1 - \mathbf{X}_2) \cdot (\gamma(v) \vec{v}, i \gamma(v) c)}{(\mathbf{X}_1 - \mathbf{X}_2) \cdot \mathbf{V}_1}.
\label{eq:60}
\end{equation}

Dividing (\ref{eq:60}) by (\ref{eq:59}) we obtain:
\begin{equation}
\frac{i c \delta \tau_1}{i c \delta \tau_1^{(0)}} = \frac{(\mathbf{X}_1 - \mathbf{X}_2) \cdot (\gamma(v) \vec{v}, i \gamma(v) c)}{(\mathbf{X}_1 - \mathbf{X}_2) \cdot (\vec{0}, i c)}.
\label{eq:61}
\end{equation}

The Coulombian interaction between point charges $Q$ and $q$ is now replaced with the Coulombian interaction between segments of length $i c \delta \tau_1$ and $i c \delta \tau_2$ on worldlines with constant linear charge densities. As a result the electric force $\frac{q Q}{r^2}$ acting on the test charge is replaced with a four-force $\frac{q Q}{r^2} \frac{i c \delta \tau_1}{i c \delta \tau_1^{(0)}}$ that depends on the velocity of the test particle. 

When the source charge is moving with uniform velocity we work in a reference frame in which the source particle is at rest. When the source charge is moving with uniform acceleration we work in a reference frame in which the test point and the center of the hyperbola are synchronous. In this reference frame $\vec{F} = \frac{q Q}{r^2} \frac{\vec{R}}{R}$. In the case of uniform velocity $R = r$, and in the case of uniform acceleration $R = MP$. We have:
\begin{equation}
\mathbf{X}_2 - \mathbf{X}_1 = ( \vec{R}, i R ), \ \ \ {\rm (retarded)}
\label{eq:62}
\end{equation}
\begin{equation}
\mathbf{X}_2 - \mathbf{X}_1 = ( \vec{R}, -i R ). \ \ \ {\rm (advanced)}
\label{eq:63}
\end{equation}

Substituting expressions (\ref{eq:62}) and (\ref{eq:63}) into equation (\ref{eq:61}) gives:
\begin{equation}
\frac{i c \delta \tau_1}{i c \delta \tau_1^{(0)}} = \gamma(v) \left( 1 - \frac{\vec{R} \cdot \vec{v}}{R c} \right), \ \ \ {\rm (retarded)}
\label{eq:64}
\end{equation}
\begin{equation}
\frac{i c \delta \tau_1}{i c \delta \tau_1^{(0)}} = \gamma(v) \left( 1 + \frac{\vec{R} \cdot \vec{v}}{R c} \right). \ \ \ {\rm (advanced)}
\label{eq:65}
\end{equation}

In a time symmetric formulation the contribution of the retarded four-force is $\frac{1}{2} \frac{q Q}{r^2} \gamma(v) \left( 1 - \frac{\vec{R} \cdot \vec{v}}{R c} \right)$, and the contribution of the advanced four-force is $\frac{1}{2} \frac{q Q}{r^2} \gamma(v) \left( 1 + \frac{\vec{R} \cdot \vec{v}}{R c} \right)$. Adding these two contributions together one obtains $\frac{q Q}{r^2} \gamma(v)$, the spatial part of the electrodynamic four-force in the direction of $\vec{R}$. 

The model is almost as successful when calculating the imaginary part of the four-force. In this case the contribution of the retarded four-force is $i \frac{1}{2} \frac{q Q}{r^2} \gamma(v) \left( 1 - \frac{\vec{R} \cdot \vec{v}}{R c} \right)$, and the contribution of the advanced four-force is $-i \frac{1}{2} \frac{q Q}{r^2} \gamma(v) \left( 1 + \frac{\vec{R} \cdot \vec{v}}{R c} \right)$. Adding these two contributions together one obtains $-i \frac{q Q}{r^2} \gamma(v) \frac{\vec{R} \cdot \vec{v}}{R c} = -i \gamma(v) \frac{\vec{F} \cdot \vec{v}}{c}$. The sign is incorrect, and we have to simply change it by an {\it ad hoc} rule. We hope that future research will elucidate this circumstance. 

One could also argue that, in the case of motion with variable acceleration, the four-force obtained from our model is not orthogonal to the four-velocity. A small deviation from orthogonality could indeed exist, and our geometrical recipe in this case amounts to supplementing the antisymmetric electromagnetic field tensor $\mathcal{F}_{ik}$ with a small (possibly symmetric) tensor $\mathcal{G}_{ik}$. We expect the contribution of $\mathcal{G}_{ik}$ to average out to zero. In any case, by requiring that the rest mass of an electron does not depend on its history, and by using Stoke's theorem, one obtains the two homogeneous Maxwell equations \cite{galeriuMax}. Other relations could also be imposed upon $\mathcal{F}_{ik}$ and $\mathcal{G}_{ik}$. Roscoe \cite{roscoe}, using the orthogonality between certain linear operator spaces, reaches the same conclusion, that the Lorentz four-force needs more than the antisymmetric electromagnetic field tensor, and identifies the additional $\mathcal{G}_{ik}$ tensor with a massive photon. The condition that Roscoe imposes, the restoration of the non-relativistic action and reaction principle, is probably at odds with the non interaction theorem. More research is needed to determine the exact physical properties of the $\mathcal{G}_{ik}$ tensor.

We will conclude this article with another interesting observation. Was Minkowski himself on the verge of discovering the possibility of replacing the interacting material point particles with infinitesimal length elements along their worldlines? This is indeed very likely, given the words "Let BC be an infinitely small element of the worldline of F; further let B* be the light point of B, C* be the light point of C on the worldline of F* [...]" that he uses when describing his first theory of gravitational interaction \cite{minkowski2}.

\vskip 30pt
\begin{eref}

\bibitem{minkowski} Hermann~Minkowski, \lq\lq Space and Time\rq\rq, published in Albert Einstein {\it et al.}, {\it The Principle of Relativity}, (Dover, New York, 1952).

\bibitem{born} Max~Born, \lq\lq Die Theorie des starren Elektrons in der Kinematik des Relativit\" atsprinzips\rq\rq,
{\it Annalen der Physik} {\bf 30}, 1 (1909).

\bibitem{sommerfeld} Arnold~Sommerfeld, \lq\lq Zur Relativit\" atstheorie. II. Vierdimensional Vektoranalysis\rq\rq, {\it Annalen der Physik} {\bf 33}, 42 (1910).

\bibitem{laue} Max~von~Laue, {\it La Th\'eorie de la Relativit\'e}, (Gauthier-Villars, Paris, 1924), tome 1, 166.

\bibitem{pauli} Wolfgang~Pauli, {\it Theory of relativity} (Dover, New York, 1981), 90.

\bibitem{walter} Scott~Walter, \lq\lq Breaking in the 4-vectors: the four-dimensional movement in gravitation, 1905 - 1910\rq\rq, published in J\"urgen Renn (ed.), {\it The Genesis of General Relativity} (Springer, Berlin, 2007), volume 3.

\bibitem{galeriu3D} C\u alin~Galeriu, \lq\lq Addition of velocities and electromagnetic interaction: geometrical derivations using 3D Minkowski diagrams\rq\rq, {\it Apeiron} {\bf 10}, 1 (2003).

\bibitem{galeriuAAD} C\u alin~Galeriu, \lq\lq Time-Symmetric Action-at-a-Distance Electrodynamics and the Structure of Space-Time\rq\rq, {\it Physics Essays} {\bf 13}, 597 (2000).

\bibitem{galeriuRR} C\u alin~Galeriu, \lq\lq Radiation Reaction 4-Force: Orthogonal or Parallel to the 4-Velocity?\rq\rq, {\it Annales de la Fondation Louis de Broglie} {\bf 28}, 49 (2003).

\bibitem{wheelerfeynman1} John~A.~Wheeler and Richard~P.~Feynman, \lq\lq Interaction with the Absorber as the Mechanism of Radiation\rq\rq, {\it Reviews of Modern Physics} {\bf 17}, 157 (1945).

\bibitem{wheelerfeynman2} John~A.~Wheeler and Richard~P.~Feynman, \lq\lq Classical Electrodynamics in Terms of Direct Interparticle Action\rq\rq, {\it Reviews of Modern Physics} {\bf 21}, 425 (1949).

\bibitem{costa2} O.~Costa de Beauregard, \lq\lq La th\'eorie de l'interaction \'electromagn\'etique de Wheeler et Feynman\rq\rq, {\it La Revue Scientifique} {\bf 3305}, 34 (1950).

\bibitem{costa3} O.~Costa de Beauregard, \lq\lq Une r\'eponse \`a l'argument dirig\'e par Einstein, Podolsky et Rosen contre l'interpr\'etation bohrienne des ph\'enom\`enes quantiques\rq\rq, {\it Comptes rendus des s\'eances de l'Acad\'emie des Sciences} {\bf 236}, 1632 (1953).

\bibitem{costa} O.~Costa de Beauregard, \lq\lq Time symmetry and interpretation of quantum mechanics\rq\rq, {\it Foundations of Physics} {\bf 6}, 539 (1976). 

\bibitem{woodward} James~F.~Woodward, \lq\lq Are the Past and the Future Really Out There?\rq\rq, {\it Annales de la Fondation Louis de Broglie} {\bf 28}, 549 (2003).

\bibitem{synge} J.~L.~Synge, {\it Relativity: The Special Theory}, (North-Holland, Amsterdam, 1958), 393.

\bibitem{fabri} Luca Fabri, \lq\lq Free falling electric charge in a static homogeneous gravitational field\rq\rq, {\it Annales de la Fondation Louis de Broglie} {\bf 30}, 87 (2005).

\bibitem{feynman} Richard~P.~Feynman, {\it Feynman Lectures on Gravitation} (Addison-Wesley, Massachusetts, 1995), 123.

\bibitem{lamont} Colin LaMont, \lq\lq Relativistic Direct Interaction Electrodynamics: Theory and Computation\rq\rq, (B.A. Thesis, Reed College, Oregon, 2011), 41.

\bibitem{galeriuMax} C\u alin~Galeriu, \lq\lq A derivation of two homogenous Maxwell equations\rq\rq, {\it Apeiron} {\bf 11}, 303 (2004).

\bibitem{roscoe} David~F.~Roscoe, \lq\lq Maxwell's Equations: New Light on Old Problems\rq\rq, published in Valeri Dvoeglazov (ed.) {\it Einstein and Poincar\'e: The Physical Vacuum} (Apeiron, Montreal, 2006).

\bibitem{minkowski2} Hermann~Minkowski, \lq\lq Die Grundgleichungen f\"r die elektromagnetischen Vorg\"ange in bewegten K\"orpern\rq\rq, {\it G\"ottinger Nachrichten}, 53 (1908).

\end{eref}

\section*{Note added on 24 February 2017}

The theory of time symmetric electrodynamic interactions presented in the first draft in 2015 (and also in 2000) has suffered an important change. Two problems have plagued the old theory. One was the incorrect sign of the imaginary component of the four-force, already discussed in the first draft (and in 2000). The second problem relates to the calculation of the magnitude of the four-force produced by an electric charge in hyperbolic motion. Let's look at the retarded component along $\overrightarrow{QP}$. Since $\overrightarrow{QP} = \overrightarrow{QM} + \overrightarrow{MP}$, we expect an electric field of magnitude $E_{MP} = \frac{1}{2} \frac{e}{r^2}$ along $\overrightarrow{MP}$. The same retarded contribution can be evaluated in the reference frame that is co-moving with the source charge at $Q$. Since $\overrightarrow{QP} = \overrightarrow{QT} + \overrightarrow{TP}$, we should be able to calculate the electric field $E_{TP}$ along $TP$  and relate it to the electric field $E_{MP}$ along $MP$. Let $V$ be the projection of point $T$ on the $Ox$ axis. Since $\overrightarrow{TP} = \overrightarrow{TV} + \overrightarrow{VM} + \overrightarrow{MP}$, and since $QM$, $QT$, $TV$, and $VM$ are restricted to the $(x,O,ict)$ plane, it follows that $E_{MP} = E_{TP} * \frac{MP}{TP}$. Since $TP = - i \rho \sin(\theta)$ and $MP = - i a \sin(\theta)$, it follows that $E_{MP} = E_{TP} * \frac{a}{\rho}$. We therefore expect an electric field of magnitude $E_{TP} = \frac{1}{2} \frac{e}{r^2} \frac{\rho}{a}$ along $\overrightarrow{TP}$. However, in the reference frame in which the test particle at $P$ and the center of the hyperbola are simultaneous, a test particle at rest of length $s_o$ will interact with a segment $S_o$ on the retarded branch of the hyperbola, whose length can be calculated as $S_o = s_o \frac{a}{\rho}$. The fraction is exactly the reciprocal of what is expected!

At this point in time it seems that our initial method of keeping the length of the test particle constant, while allowing for a change in the length of the source particle, has to be abandoned. The interacting particles will still have infinitesimal length segments whose end points are connected by light signals, but this time the length of the source particle will be held constant, while the length of the test particle will be allowed to vary. A similar approach was suggested by Kevin Brown in Physics in Space and Time. What is calculated in this way is no longer a force, but a force density (in better agreement with the theory described in 2000), and we will add the retarded and advanced contributions as initially described. This new approach solves both problems outlined above. A more detailed description of the updated theory will be given at a later date.

\end{document}